\documentstyle[floats,twocolumn,prl,aps]{revtex}
\begin{document}
\draft
\def\ref{\par\noindent\hangindent=3mm\hangafter=1}
\narrowtext
{
\title{
Small-$\bf{q}$ electron-phonon scattering and linear DC resistivity
in high-$T_c$ oxides}
}
\author{G. Varelogiannis and E.N. Economou}
\address{
Institute of Electronic Structure and Laser, FORTH,
P.O. Box 1527,
Heraklion Crete 71110, Greece}
\author{\parbox{397pt}{\vglue 0.3cm \small
We examine the effect on the DC resistivity
of small-$\bf{q}$ electron-phonon scattering, in a system with the electronic
topology of the high-$T_c$ oxides.
Despite the fact that the scattering is
dominantly forward,
its contribution to the transport can be significant due to
``ondulations'' of the bands in the flat region and to the umpklapp process.
When the extended van-Hove singularities are
sufficiently close to $E_F$
the acoustic branch of the phonons contribute significantly to the transport.
In that case one can obtain linear $T$ dependent resistivity down to
temperatures as low as 10 $K$, even if electrons are scattered also by
optical phonons of about $500 K$ as reported by Raman measurements.
}}
\maketitle
\par
One of the most puzzling experimental facts 
in the optimally doped oxides is the linear
temperature dependence of their $DC$ resistivity for
very low temperatures.
In $Bi_2(Sr_{0.97}Pr_{0.03})_2CuO_6$ the $DC$ resistivity is linear
from $10K$ up to several hundreds of Kelvins \cite{King}. 
This has been considered as a sign of absence of electron phonon scattering.
In fact, eventhough
a linear $T$ dependence of the DC resistivity is also possible in the case of
electron-phonon scattering, this happens for 
temperatures that are higher than about 
one fourth of the characteristic phonon energies.
From this perspective, the data of Ref. \cite{King}  
indicate a complete absence of electron-phonon scattering 
for frequencies above $50 K$, while the optical phonons in the
cuprates extend
up to about $1000 K$.

To explain the linear $T$-dependence of the DC resistivity, several scenarios 
avoiding electron-phonon scattering have been proposed.
A first approach was the so called Marginal Fermi Liquid \cite{MFL}, where 
the linearity was evidence of deviation from the classical Fermi liquid
behavior in the case of 
some special collective excitation scattering. A second type of approach
was that of gauge theory models where the linearity is a signature of a complete
breakdown of Landau Fermi Liquid theory replaced by a quasi-one dimensional
Luttinger liquid behavior \cite{gauge}. A third type of approach
was that of a nearly antiferomagnetic Fermi liquid \cite{paramagnons}, in which
case there is singular electron-paramagnon scattering supposed to be
also at the origin
of
d-wave superconductivity. Finally a linear $T$-dependent resistivity has also been
associated with the presence of van Hove singularities in the vicinity of the
Fermi surface and simple Coulomb scattering \cite{vanHove}. 
The relevance of the last 
two approaches has been questioned 
recently by Hlubina and Rice \cite{Hlubina}. 

Although
in all the previous scenarios electron-phonon scattering is
neglected, there is strong evidence from Raman experiments that there
is significant coupling of the electrons with
the optical oxygen vibrations \cite{Zech}, and the absence of signature
of this coupling in the DC
resistivity is an unresolved puzzle. 
It also appears rather unphysical that phonons could be 
irrelevant for the $T$-dependence of the DC resistivity at so high 
temperatures.

We will propose here an alternative scenario that could reconcile the Raman and
DC transport results. In this scenario the electron-phonon scattering
is the dominant $T$-dependent resistive mechanism, the phonon spectrum
and electronic topology are similar to that 
of the oxides yet the resistivity could be linear down to temperatures as
low as $10 K$ in optimally doped materials. 
A basic assumption of our approach is the dominance of small momentum transfer
process in the electron-phonon scattering.
This assumption, although it has not been based on solid physical arguments
or computations, nevertheless has been succesfull in interpreting many
puzzling features of the high-$T_c$ superconductors.
Some features are
the peak in the microwave conductivity \cite{VPrc},
the momentum dependence of the anomalous dip above the gap
and the enhancement of the anisotropy close to $T_c$ reported by ARPES on
$BiSr_2CaCu_2O_8$ 
\cite{DOSdriven},
as well as the presence of different gap symmetries in different
oxides and even variations of the gap symmetry with doping \cite{SDwave}.
An effectively 
similar hypothesis of small-$q$ scattering in the oxides is actually 
investigated by many authors in different contexts 
\cite{Chakravarty}. The dominance of forward scattering in the electron-phonon
interaction could result from strong Coulomb correlations of
the carriers \cite{Zeyher} that may brink the electronic system in the 
vicinity of a phase separation instability \cite{PS,Andr}.           

One can briefly sketch our approach as follows. When the electron-phonon
scattering is limited to small momentum transfer processes, then the 
contribution of the
acoustic branch of the phonons is energetically separated from
that of the 
optical branches. The acoustic branch extends up to an energy of the
order $\Omega_A=q_c\upsilon_s$ where $\upsilon_s$ is the sound velocity
and $q_c$ a characteristic momentum cut-off of the scattering.
Analyzing the phenomenology of $BiSr_2CaCu_2O_8$ 
we obtained $q_c\approx k_F/10$ \cite{DOSdriven,SDwave}.
Taking $\upsilon_s\approx 10^{-2}\upsilon_F$ and an average Fermi velocity
$\upsilon_F\approx 2\times 10^7 cm s^{-1}$
in agreement with infrared measurements \cite{IRv} 
we obtain $\Omega_A\approx 40 K$.
The energetic cut-off for the acoustic branch,
which is the effective ``transport'' Debye frequency, being not larger than
$50 K$, the
resistivity can be linear for temperatures as small as $10 K$.
The optical phonons, for which there is evidence from Raman that
contribute dominantly to the electron-phonon scattering
in the region of frequencies between
$300$ and $600 $ Kelvins, in principle should
also dominate the transport. However there is a significant difference 
between dominantly forward scattering of electrons 
with optical and acoustic phonons. 
In the first case
we exchange small-momenta but large 
energies while in the second case we exchange
small momenta and small energies and the phase space for these two types of
processes is very different depending on how close the
flat band regions are to the Fermi surface. 
In fact in order to have contribution
from small momentum transfer processes to the transport, and avoid the
well known 
$1-\cos\theta$ coefficient from the 
Boltzmann equation, it is necessary to scatter
between points with opposite Fermi velocities. 
In the case of small momentum transfer,
this happens principally because of the 
``ondulations'' of the bands in the flat region, and also
because of the umpklapp processes. In both cases the flat regions are concerned,
and depending on the distance of these regions from the Fermi surface, the 
acoustic branches can dominantly
contribute to the transport. When the acoustic branch
gives a dominant contribution,  
the resistivity is linear down to very low temperatures
since the effective Debye frequency is the cut-off of the acoustic branch
not larger than few tenths of Kelvins if the scattering
is dominantly forward. 

In order to illustrate how realistic are the previous arguments   
for the high-$T_c$'s,
we consider a tight binding  
fit to the ARPES reported Fermi surface
and dispersion of $BiSr_2CaCu_2O_8$. 
The most characteristic feature
is the presence of extended van Hove 
singularities that cover about $30 \%$ of the
Brillouin zone, and in order 
to produce such extended van Hove singularities
in tight binding,
hoping terms up to the fifth nearest neighbors
were found to be necessary: 
$$
E_{\bf{k}}= t_1(\cos k_x+\cos k_y)+t_2 \cos k_x \cos k_y +
$$
$$
t_3 {1\over 2} (\cos 2 k_x + \cos 2 k_y) +
t_4 {1\over 2} (\cos 2k_x \cos k_y + \cos k_x \cos 2k_y)
$$
$$
+t_5 \cos 2k_x\cos 2k_y 
\eqno(1)
$$
The dispersion we
consider is not very different from that considered in Ref. \cite{Radtke}
and corresponds to the set of parameters: $t_1=-0.525$, $t_2=0.0337$,
$t_3=0.0287$, $t_4=-0.175$ and $t_5=0.0175$. This dispersion fits well
the ARPES results \cite{Shen} 
and especially the extended van Hove singularities
taken exactly at $E_F$.               
We show in figure (1a) a quarter of the Brillouin zone 
and the corresponding electron
dispersion given by (1). 
The white area is a region of $50 K$ around the Fermi surface.
One can see that when the extended van Hove singularities are sufficiently 
close to the Fermi surface, the ondulations of the band in the 
flat region creates effective branches 
of Fermi surface around the points ($\pi,0$) and ($0,\pi$) in addition
to the principle branch around ($\pi,\pi$). The scattering from one branch
to the other is associated with a 
reversal of the Fermi velocity. These ondulations together 
with the umpklapp processes imply that even small momentum transfer processes
give a significant contribution to the transport.
Although the flat band fluctuates over few tenths of $meV$ 
the electronic density of states
has a very sharp peak characteristic of extended saddle points.
We show in Figure 2 the DOS for the dispersion shown in figure 1a where
the van Hove peak lies exactly at $E_F$. 

The phase space available for transport efficient electron phonon scattering 
can be measured considering the following definition
of the transport Eliashberg function 
$$
\alpha_{tr}^2F_{tr}(\Omega)\approx \sum_{\bf{k},\bf{k'}}
{\tilde{A}_{\bf{k}\bf{k'}}\over 2} 
{(\bf{\upsilon_k}-\bf{\upsilon_{k'}})^2\over
|\bf{\upsilon_k}||\bf{\upsilon_{k'}}|}
\delta
(\Omega_{\bf{k}-\bf{k'}}-E_{\bf{k}}+E_{\bf{k'}})
\eqno(2)
$$
where an electron scatters from the occupied state $E_{\bf{k}}$ to the 
empty state $E_{\bf{k'}}$. The velocities are defined by
$\bf{\upsilon_k}=\bf{\nabla_k} E_{\bf{k}}$ 
and in the case of elastic scattering
in an isotropic system we have ${(\bf{\upsilon_k}-
\bf{\upsilon_{k'}})^2\over 2
|\bf{\upsilon_k}||\bf{\upsilon_{k'}}|}\approx 1-\cos\theta$
where $\theta$ is the angle between  the velocities
$\bf{\upsilon_{\bf{k}}}$ and $\bf{\upsilon_{\bf{k'}}}$.
The coefficients $\tilde{A}_{\bf{k}\bf{k'}}$ are scattering amplitude matrix
elements which are too complicated for explicit evaluation. Thus we shall
use the following arbitrary but simple analytical form, which satisfies
our basic requirement that $\tilde{A}_{\bf{k}\bf{k'}}$ should become
negligible for $|\vec{k}-\vec{k'}|>q_c$:
$$
\tilde{A}_{\bf{k}\bf{k'}}\approx
-\tilde{g}_{A,O}^2 \biggl( 1 + 2{2-\cos (k_x-k'_x)-\cos (k_y-k'_y)\over q_c^2}
\biggr)^{-1}
\eqno(3)
$$
where $\tilde{g}_{A}^2\approx g_A^2\sqrt{2-\cos(k_x-k'_x)-\cos (k_y-k'_y)}$
is proportional to the scattering cross section with the acoustic and
$\tilde{g}_{O}^2\approx g_O^2$ to the scattering cross section with
the optical phonons. The scattering is therefore dominated by the
processes in which the exchanged momentum is smaller than $q_c$ that 
plays the role of a smooth momentum cut-off. The exact form of the
momentum cut-off is irrelevant for our qualitative arguments.

In reality,
the scattering amplitude coefficients are also 
frequency dependent especially for very 
anisotropic systems as the high-$T_c$ oxides. In fact the phonon system
in the oxides is 
three dimensional but the electronic system is two dimensional, and obviously 
all phonons will not have the same probability to scatter with electrons.
Phonon symmetry considerations also influence the probability 
of scattering. General
calculations of such probabilities 
is a rather complex task \cite{Pickett}, that
we will avoid here focusing on our phase space arguments.
We therefore adopt ad-hoc a frequency structure of the scattering
that agrees with experiment.                
In fact the optical vibrations that are relevant to the scattering by
the in plane oxygens 
are now well documented \cite{Liu} and range between 25 and 50 meV.
On the other hand there is evidence from
Raman spectroscopy \cite{Zech},
as well from a comparison to a study of
the spectral dependence of the gap ratio \cite{aspec}, that 
electrons couple strongly to these optical phonons.
This spectral structure also explains 
the disagreement between the infrared and the 
other gap measurements \cite{aspec,VPrc}.
We therefore consider a spectrum consisting by the acoustic
branch that extends up to $\Omega_A\approx 50 K$ 
and the optical branches that
we take as a constant distribution between $\Omega_1\approx 25meV$
and $\Omega_2\approx 50meV$.

To obtain the transport parameters we follow the conventional approach and
we suppose that to a first approximation, 
the transport scattering time
has the same definition as the quasiparticle lifetime except that 
the normal Eliashberg function is replaced by
the transport Eliashberg function.
This assumption is common to many theoretical approaches 
to the linear resistivity problem \cite{MFL,gauge,paramagnons,vanHove}.
For the quasiparticle lifetime
$1/\tau=-2 Im(\Sigma)$ only the lowest order contribution
of phonons in the electronic self energy $\Sigma$ is taken in agreement
with Migdal's theorem \cite{Mahan} leading to 
$$
{1 \over \tau}= \pi
\int d\Omega \alpha_{tr}^2 F_{tr} (\Omega)
\biggl[ 2\coth {\Omega\over 2T} - 
$$
$$\tanh {\omega+\Omega\over 2T}
+\tanh {\omega-\Omega\over 2T}\biggr]
\eqno(4)
$$
Certainly,                                                                     
for a more accurate approach we should solve the Kubo problem for the 
specific anisotropic situation that we consider, but such treatment is 
beyond the scope of this manuscript.
We focus here on the DC 
resistivity which corresponds to the $\omega\rightarrow
0$ limit.
The transport scattering time for our spectrum 
is given by
$$
{1 \over \tau}\approx 8\pi G_A\int_0^{\Omega_A}
d\Omega {\Omega^2 \over \exp (\beta\Omega)-\exp (-\beta\Omega)}+
$$
$$
+
4\pi G_O T \biggl[ {1\over 2 }\ln { 0.5 [\exp (\beta \Omega_2)+
\exp(- \beta \Omega_2)]-1\over
0.5 [\exp (\beta \Omega_2)+
\exp(- \beta \Omega_2)]+1}-
$$
$$
-
{1\over 2 }\ln { 0.5 [\exp (\beta \Omega_1)+
\exp(- \beta \Omega_1)]-1\over
0.5 [\exp (\beta \Omega_1)+
\exp(- \beta \Omega_1)]+1}\biggr]
\eqno(5)
$$.

The coefficients $G_A$ and $G_O$ are obtained from equation (2).
It is important to notice that for the van Hove singularity sufficiently
close to $E_F$, both coefficients are significant despite the fact that 
the scattering is dominantly forward. We show in figure 2 the evolution of the
ratio $G_A/G_O$ as a function of the position of the van Hove
singularity with respect to the Fermi level when $g_A\approx g_O$.
ARPES measurements in $BiSr_2CaCu_2O_8$ report the van
Hove singularities at maximum few tenths of meV below $E_F$, and as we see in
figure 2, in that case the acoustic branch gives a non negligible
contribution to the transport.
It is remarkable that when the singularity is pushed 
sufficiently
close to $E_F$ we obtain $G_A > G_O$.  
In this regime the resistivity that is proportional to $1/\tau$ is 
expected to be linear down to very low temperatures.
Indeed we plot in figure 3
the resistivity in arbitrary units as a function of temperature
when the van Hove peak in the electronic DOS lies about $30 K$ below 
the Fermi level.
We see the remarkable linearity from $10 K$
up to $900 K$ that reproduces qualitatively the results of Ref. \cite{King}.

Linear $T$ dependence of the resistivity do not necessarily
means $G_A > G_O$. In fact, 
the relative influence of the acoustic branch to the 
slope of the $T$ dependence of the resistivity is much larger than indicated
from the ratio $G_A/G_O$. One can understand this taking for example the 
high temperature expansion of equation (5)
$ 
1/ \tau\approx 
2\pi T
\bigl[G_A \Omega_A^2 + G_O \ln (\Omega_2/\Omega_1)\bigr]
$.
Both acoustic and optical phonons give a linear $T$ dependence in this
regime. However the contribution of the optical branches to the slope
depends logarithmically on the width of the optical part of the spectrum,
while the acoustic branch contributes as the square of the
number of Kelvins to which $\Omega_A$ corresponds.
For our rather concentrated optical spectrum 
the acoustic branch determines the $T$-slope of the resistivity 
even for $G_A < G_O$. Therefore, if we have a linear $T$ dependence
in the region $0.2\Omega_A<T<\Omega_1$ because of the acoustic
branch, when at higher temperatures
we enter the regime $T>\Omega_2$, contrary to what one might naively expect,
there is {\it not a significant variation of the slope}
corresponding to the entry in game of the optical phonons even if the
electrons are strongly coupled to them. 
Notice also that concerning the
pairing, the optical phonons could be dominant within our approach.
However,
since they are inefficient for
the transport, the total coupling strength for superconductivity
can be an order of magnitude higher
than the effective total transport coupling strength, in agreement
with the phenomenology of high-$T_c$'s.

In conclusion,
a small momentum cut-off of the electron-phonon
scattering, first limits the acoustic branch to energies of few tenths
of Kelvins and secondly for systems with the electronic topology of the
oxides implies a significant relative contribution of the acoustic phonons 
to the
transport. In that case the effectively relevant 
Debye energy for the transport
is the cut-off of the acoustic branch and the resistivity can be linear
for temperatures as small as $10K$ eventhough optical Raman active
phonons of $\approx 500K$ couple strongly to electrons
and could even be responsible for superconductivity.

Numerous stimulating discussions with G.C. Psaltakis are
gratefully acknowledged.


{\Large \bf Figure Captions}

{\bf Figure 1:} (a) The dispersion of Equation 1 on a quarter of the
Brillouin zone in $\pi/100$ units. 
The white region represents an energy window of
$50 K$ around the Fermi level. The extended van Hove singularities 
correspond to the 
plateaus centered at $(0,100)$ and $(100,0)$ ($\bar{M}$ points in 
$BiSr_2CaCu_2O_8$) which cover about one third ofthe Brillouin zone.
(b) The corresponding electronic density of states (DOS) with the
sharp peak at the Fermi level indicating the extended van Hove
singularity.

{\bf Figure 2:} The ratio $G_A/G_O$ as a function of the 
energetic
position of the van Hove peak in the DOS with respect to the Fermi level.

{\bf Figure 3:} The resistivity in arbitrary units as a function
of temperature when the van Hove peak lies about $30 K$ below $E_F$
for the spectrum described in the text where optical phonons 
in the range $25-50 meV$ couple strongly to electrons. The linear
behavior from $10K$ up to $900 K$ reproduces qualitatively the 
results of \cite{King}.

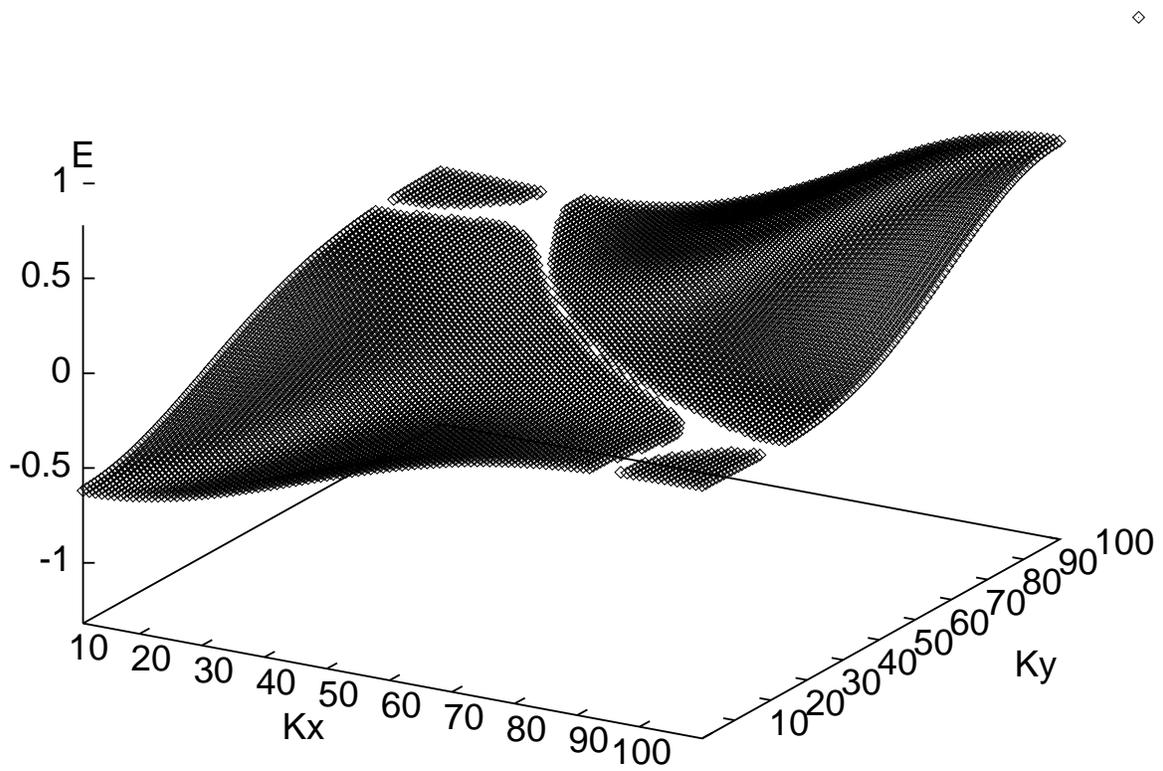
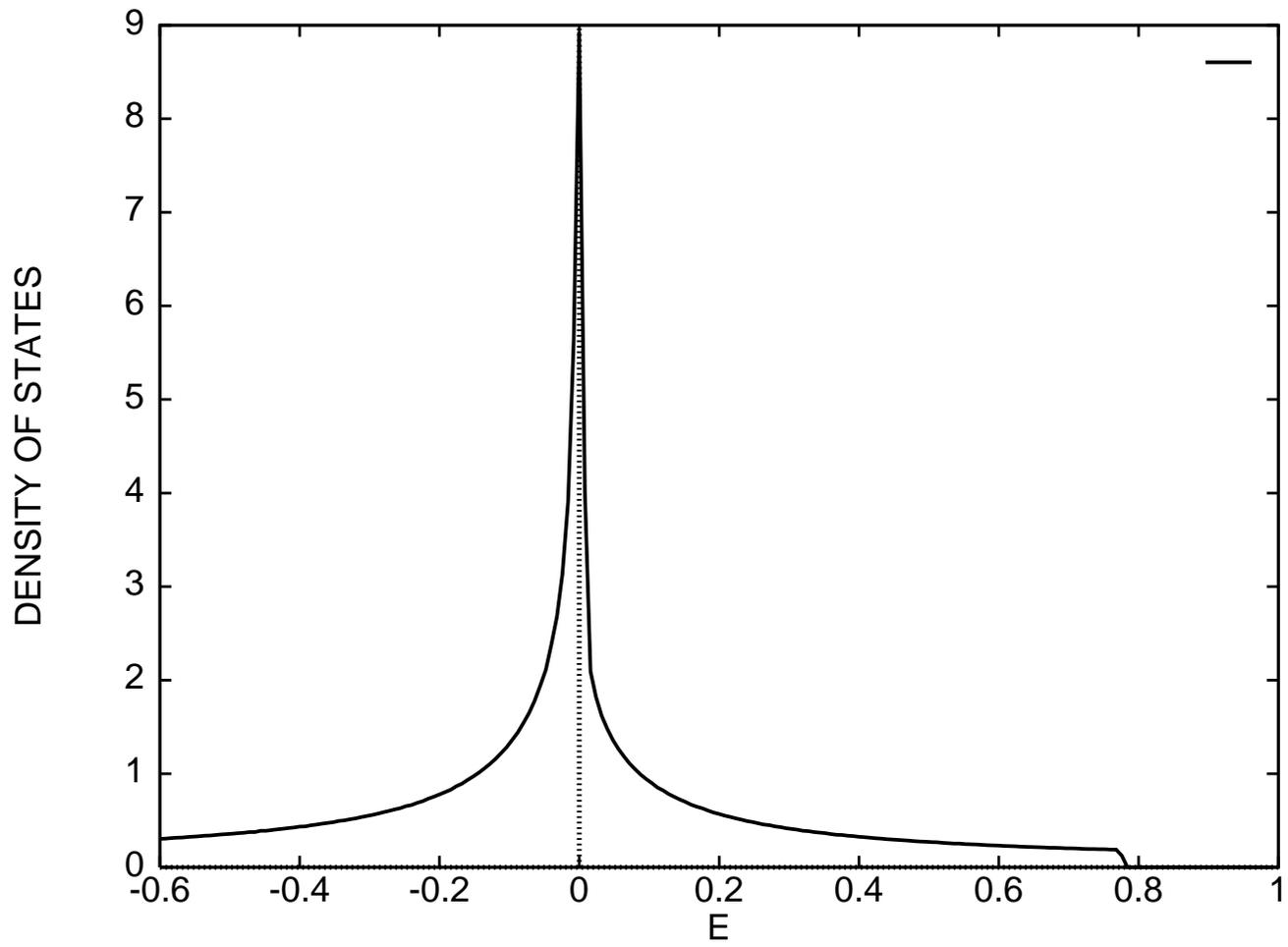
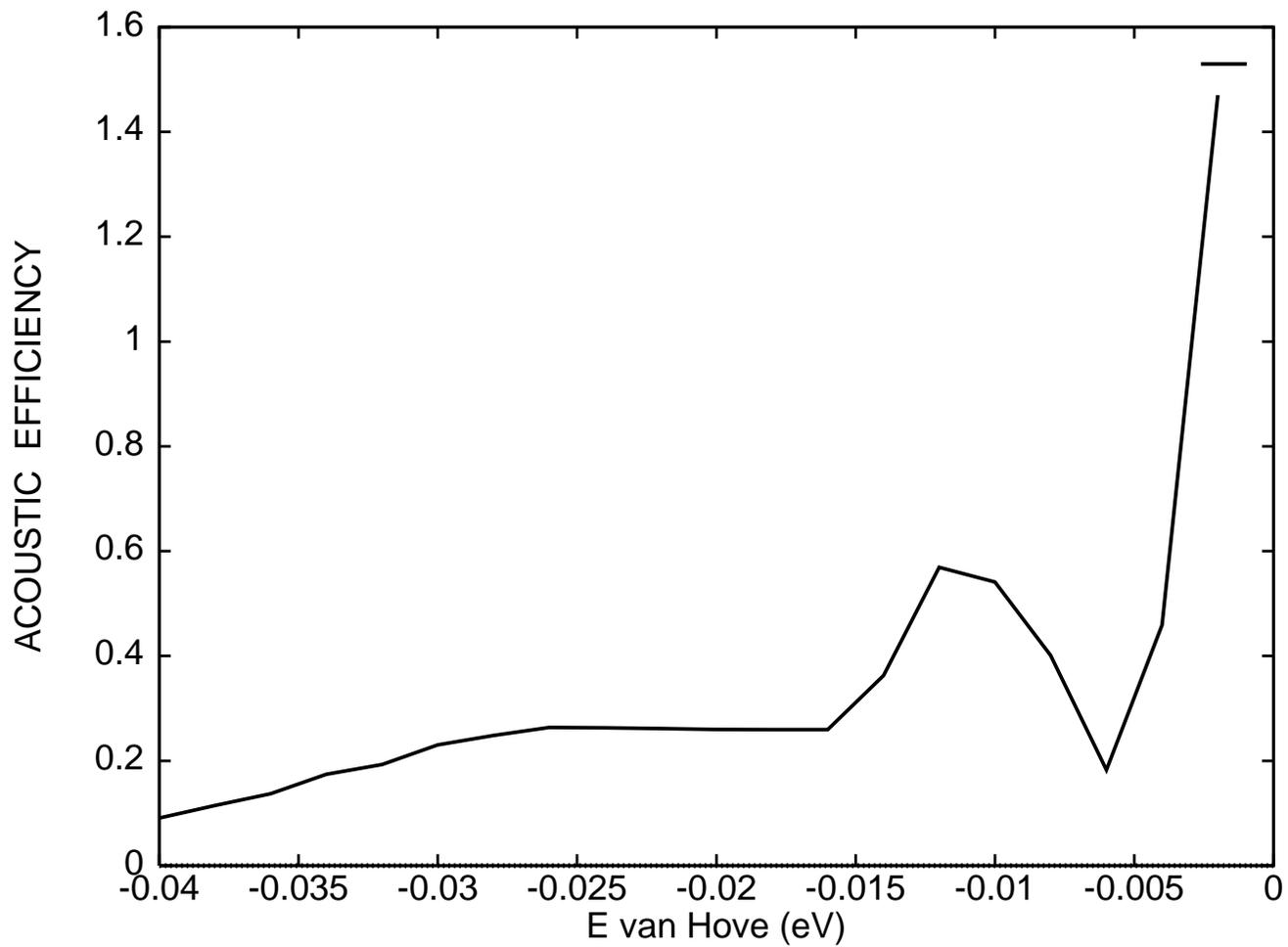
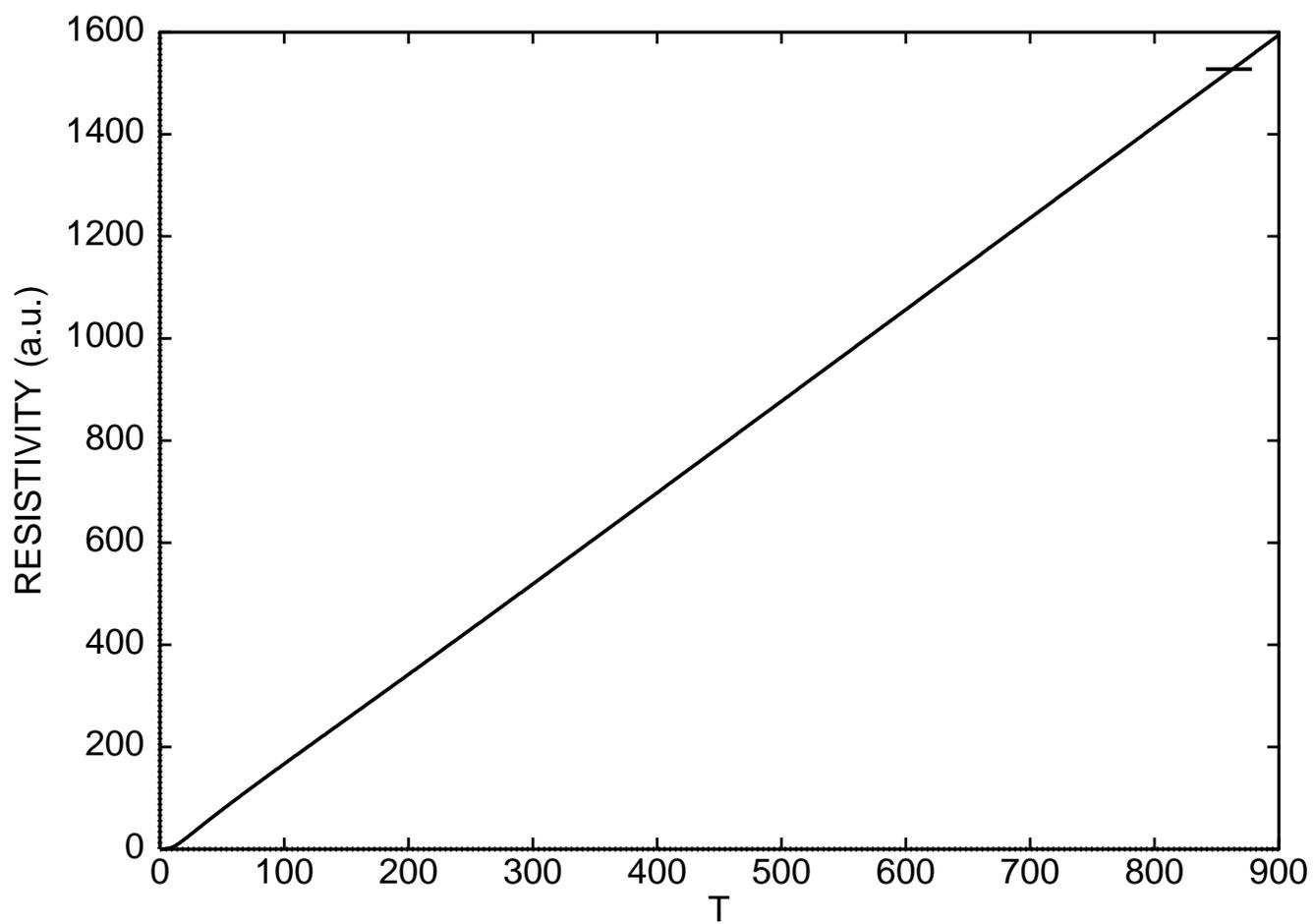
\end{document}